\documentclass[pra,letter
,showpacs,preprintnumbers,floatfix,superscriptaddress]{revtex4}
\usepackage{amsmath}
\usepackage{amsfonts}
\usepackage{amssymb}
\usepackage{graphicx}
\usepackage{epsfig}
\usepackage{dcolumn}
\usepackage{bm}
\usepackage{relsize}
\usepackage[usenames,dvipsnames]{color}	 
\usepackage[colorlinks=true,breaklinks=true,bookmarksopen=true,bookmarksnumbered=true,citecolor=blue,linkcolor=red,hyperfootnotes=false]{hyperref}  
\usepackage{soul}
\begin{document}
\title{Elastic electron scattering from Zn, Cd, and Hg}
\author{Mehrdad Adibzadeh}
\email{madibzadeh@uwf.edu}
\affiliation{Department of Physics, University of West Florida, Pensacola, FL 32514, USA}
\author{Constantine E. Theodosiou}
\email[Corresponding author: ]{ constant.theodosiou@manhattan.edu}
\affiliation{Department of Mathematics and Physics, Manhattan University, Riverdale, NY 10471, USA.}
\date{\today}

\begin{abstract}
We present an extensive set of theoretical results for differential, integrated, and momentum transfer cross sections for the elastic scattering of electrons by zinc, cadmium, and mercury. This study extends the application of our method of calculations, previously employed for stable inert gases and alkaline-earth-metals. Our approach is a self-consistent calculation, with a semi-empirical element in the adjustable cut-off radius of the polarization potential.  Our method is expected to provide a set of accurate data for Zn, Cd and Hg, based on the satisfactory agreement in our previous investigations with experimental values and other precise theoretical results.
\end{abstract}
\maketitle

\section{Introduction}
The study of elastic scattering of electrons off atoms provides the fundamental understanding of the interaction and the atomic structure and even that of the matter in bulk. However, aside from fundamental reasons, these studies provide, in particular, data for applications.

In the case of heavy metal atoms, electron scattering has been an  important component of investigations involving plasmas in both industry and research. Mercury has drawn the most interest from experiment and theory in the past hundred years not only due to its large atomic number and the importance of relativistic effects in such an atom, but for its application in the development of fluorescent and discharge lamps. Similarly, zinc has also drawn attention to itself in the past fifteen years as a replacement for mercury in high intensity discharge lamps to limit some negative environmental impacts of mercury in such lamps.

In our previous works, where we had extended our semi empirical approach for closed shell inert gas atoms~\cite{Adibzadeh2005} to stable alkaline earth metals~\cite{Adibzadeh2004,Adibzadeh2024}, which all represent quasi-two-electron atoms.  Hence it was reasonable to test our approach for other quasi-two-electron atoms, i.e. zinc (Zn), cadmium (Cd) and mercury (Hg).

In the case of zinc, the experimental data on cross sections for elastic collision of electrons are fairly limited. Dating back to the 1930s, they include works of Brode~\cite{Brode1930}, Childs and Massey~\cite{Childs1933b}, Williams and Bozinis~\cite{Williams1975}, Burrow et al.~\cite{Burrow1976} and Marinkovi\'{c} et al.~\cite{Marinkovic2019}.

Theoretical works providing elastic cross sections for zinc were limited to three in the early 1990s. They are those of McGarrah et al.~\cite{McGarrah1991}, Yuan and Zhang~\cite{Yuan1992} and Kumar et al.~\cite{Kumar1994}. After the proposal for zinc’s possible application in discharge lamps~\cite{Born2001}, theoretical interest in cross sections for zinc increased over the past two decades comprising the works of White et al.~\cite{White2004}, Fursa et al.~\cite{Fursa2005}, Zatsarinny and Bartschat~\cite{Zatsarinny2005}, Bostock et al.~\cite{Bostock2012}, McEachran et al.~\cite{McEachran2020}, and Arretche et al.~\cite{Arretche2022}.

Experimental measurements of cross sections for cadmium also date back to the 1930s. They include the works of Brode~\cite{Brode1930}, Childs et al.~\cite{Childs1933b}, Burrow et al.~\cite{Burrow1976}, Nogueira et al.~\cite{Nogueira1987}, Marinkovi\'{c} et al.~\cite{Marinkovic1991}, Kontros et al.~\cite{Kontros2002,Kontros2003}, Sullivan et al.~\cite{Sullivan2003}, and recently of Marinkovi\'{c} et al.~\cite{Marinkovic2023}.

The theoretical calculations of cross sections for elastic scattering of electrons off cadmium atom are those of Pangantiwar and Srivastava~\cite{Pangantiwar1989}, McGarrah et al.~\cite{McGarrah1991}, Nahar~\cite{Nahar1991}, Madison et al.~\cite{Madison1991}, Yuan and Zhang~\cite{Yuan1992}, Berrington et al.~ \cite{Berrington2012}, Haque et al.~\cite{Haque2021}, Arretche et al.~\cite{Arretche2022}, and Marinkovi\'{c} et al.~\cite{Marinkovic2023}.

For mercury the experimental studies of cross sections for electron scattering began a hundred years ago. The body of experimental literature consists of the works of Maxwell~\cite{Maxwell1926}, Beuthe~\cite{Beuthe1928}, Jones~\cite{Jones1928}, Brode~\cite{Brode1929}, Arnot~\cite{Arnot1931}, Palmer~\cite{Palmer1931}, Kessler and Lindner~\cite{Kessler1965}, Deichsel et al.~\cite{Deichsel1966}, Bromberg~\cite{Bromberg1969}, Gronemeier~\cite{Gronemeier1970}, D\"{u}weke et al.~\cite{Duweke1976}, Burrow et al.~\cite{Burrow1976}, Jost and Ohnemus~\cite{Jost1979}, Hanne et al.~\cite{Hanne1980}, Holtkamp et al.~\cite{Holtkamp1987}, Peitzmann and Kessler~\cite{Peitzmann1990}, Panajotovi\'{c} et al.~\cite{Panajotovic1993}, and Zubek et al.~\cite{Zubek1995}.

The theoretical calculations of cross sections for electron scattering from mercury include the works of McCutchen~\cite{McCutchen1958}, Fink and Yates~\cite{Fink1969}, Rockwood~\cite{Rockwood1973}, Walker~\cite{Walker1975}, Lam~\cite{Lam1980}, Elford~\cite{Elford1980}, McEachran and Stauffer~\cite{McEachran1987}, England and Elford~\cite{England1991}, Sienkiewicz~\cite{Sienkiewicz1997}, Fursa et al.~\cite{Fursa2003a}, Fursa and Bray~\cite{Fursa2003b}, McEachran and Elford~\cite{McEachran2003}, Zatsarinny and Bartschat~\cite{Zatsarinny2009}, Bostock et al.~\cite{Bostock2010}, and Haque et al.~\cite{Haque2019,Haque2021}.

\section{Brief Review of the Theoretical and Computational Approach}
In the present work we follow the same method of calculations, described in our previous papers~\cite{Adibzadeh2004, Adibzadeh2005, Adibzadeh2024}. To recap, we performed the standard method of partial-wave expansion in potential scattering, where the phase shifts were obtained by solving the stationary Dirac equation. In the current work, the choices for central static atomic, exchange, and polarization potential in Dirac Hamiltonian are the same as those of Refs.~\cite{Adibzadeh2004, Adibzadeh2024}. 

Such a combination for potentials were obtained through an exhaustive analysis and comparison with the collection of all appropriate experimental and theoretical results on elastic electron scattering by a number of closed shell atoms: namely inert gases and alkaline-earth-metals. The result was a combination of the central static atomic, exchange, and polarization potentials that produced a consistent agreement between the calculated cross sections and reliable experimental and theoretical results at different collision energies.  

In this work  we did not simply assume those choices, but applied the same meticulous methodology to Zn, Cd and Hg to confirm the choices and to obtain the only free parameter in this work: the cutoff radius of the polarization potential. Through this analysis, the most consistent combination of potentials for Zn, Cd, and Hg was determined to be the Dirac-Slater atomic potential, the semi-classical exchange potential expression by Furness and McCarthy~\cite{Furness1973} and the Buckingham-type II polarization potential in the form
\begin{equation}
V_P(r) =  - \frac{\alpha _d}{2(r^2 + r_{c}^2 )^2}\, .
\end{equation}
In the above equation, $ \alpha _d $ and $r_c $ are the static atomic polarizability and the cutoff radius, respectively. The static polarizabilities, used in polarization potential, were taken from the theoretical values of Kolb et al.~\cite{Kolb1982}. 

Similar to our previous works, to determine the cutoff radius, we required $r_c$ to be a smooth, continuous and finite function of the scattered electron energy $E$. We further confirmed the functional behavior of this energy-dependent cutoff radius to be similar to those we used for alkaline-earth-metal atoms ~\cite{Adibzadeh2004,Adibzadeh2024} through comparisons with dependable theoretical and experimental data. The cutoff radius for zinc was determined to be (in atomic units)

\begin{equation}
r_c (E) = \left\{ {\begin{array}{*{20}c}
   {\frac{1}{3}\ln (\frac{E}{\mathcal{R}}) + \left\langle r \right\rangle _{4s} }
   & {E \ge 45{\rm{ eV}}},  \\\\
   {{\rm{3}}{\rm{.85}}} & {E < 45{\rm{ eV}}},  \\
\end{array}} \right.
\end{equation}

for cadmium,

\begin{equation}
r_c (E) = \left\{ {\begin{array}{*{20}c}
   {\frac{1}{3}\ln (\frac{E}{\mathcal{R}}) + \left\langle r \right\rangle _{5s} }
   & {E \ge 30{\rm{ eV}}},  \\\\
   {{\rm{4}}{\rm{.0}}} & {E < 30{\rm{ eV}}},  \\
\end{array}} \right.
\end{equation}

and for mercury,

\begin{equation}
r_c (E) = \left\{ {\begin{array}{*{20}c}
   {\frac{1}{3}\ln (\frac{E}{\mathcal{R}}) + \left\langle r \right\rangle _{6s} }
   & {E \ge 20{\rm{ eV}}},  \\\\
   {{\rm{3}}{\rm{.6}}} & {E < 20{\rm{ eV}}}.  \\
\end{array}} \right.
\end{equation}

Here $E$ is the energy of the incident electron in eV, $\mathcal{R}$ is the Rydberg 
constant ($\mathcal{R}$ = 13.605 691 72 eV), and $\left\langle r \right\rangle
_{4s} = \text{2.680 }a_0$,  $\left\langle r \right\rangle
_{5s} = \text{2.978 }a_0 $ and  $\left\langle r \right\rangle
_{6s} = \text{3.045 }a_0 $ are the expectation values of zinc's $4s$ shell, cadmium's $5s$ shell, and mercury's $6s$ shell radii, respectively. The cutoff radii for low energies are set to constant values to avoid the anomaly caused by the logarithmic term. For more on this constant value and its behavior below an energy threshold, the reader may consult with Ref.~\cite{Adibzadeh2004}. Additionally, to obtain the low energy constant values for $r_c$, comparisons with accurate theoretical data for integrated cross section were also used as guidance. Similar to our previous works, we used our modified version of the code by Salvat et al.~\cite{Salva1995} in our calculations and throughout this work and for all considered energies, we used up to 150 partial-wave phase shifts.

\section{Results and Discussion}
Our objective in this work was to compare our calculations with all available experimental measurements to date. To maintain graph readability, when comparing our values with other works, we will limit the comparisons to most recent and reliable data whenever possible. We also avoid the placement of error bars on experimental data if the uncertainty may be encompassed through an appropriate size of the marker.

\subsection{Zinc}
\begin{figure*}
\includegraphics[scale=0.92]{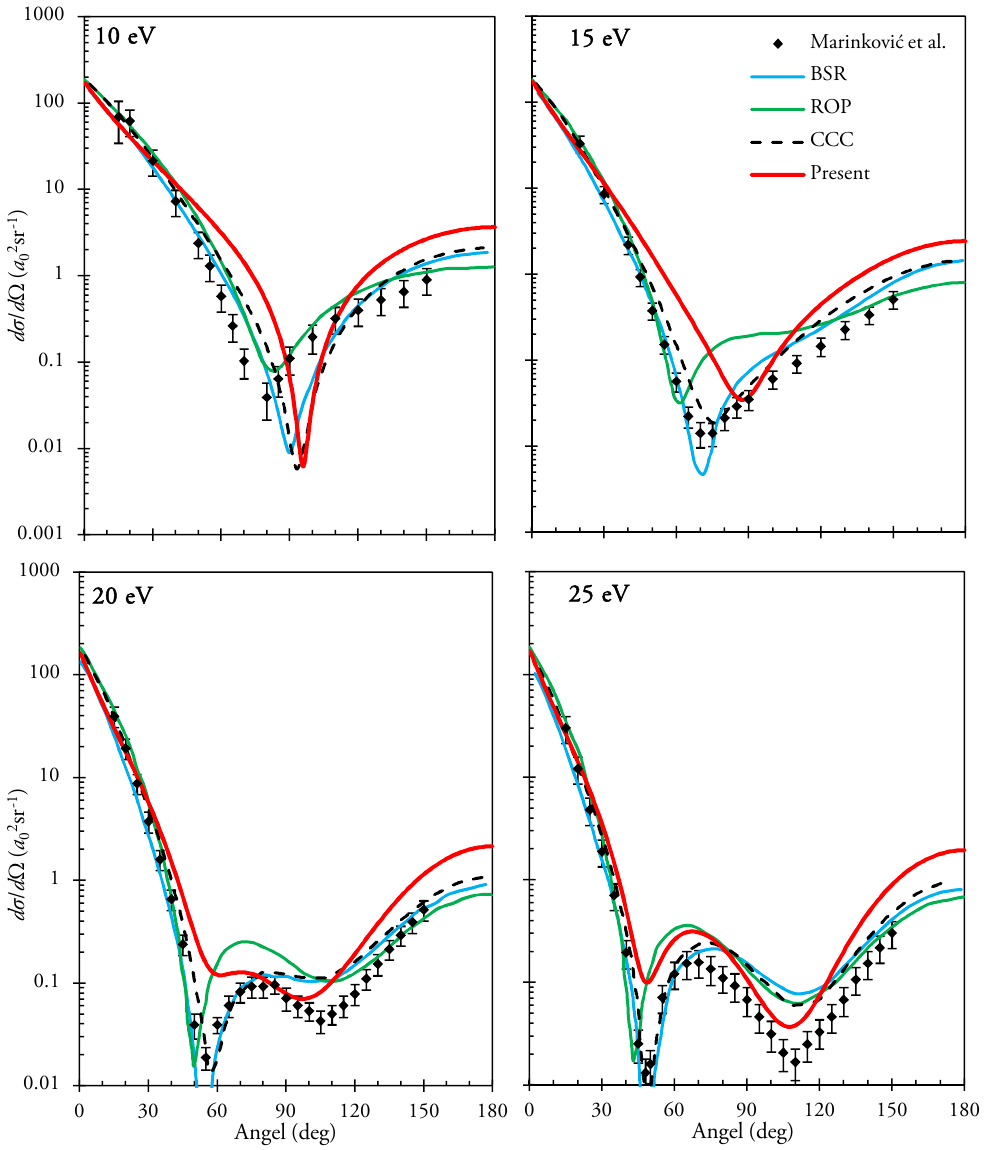}
\caption{\label{fig:zn1} Differential cross sections for elastic electron scattering from zinc at 10, 15, 20 and 25 eV: The legend in the figure describes markers for Present work; Experiment: Marinkovi\'{c} et al.~\cite{Marinkovic2019}; Other theoretical: BSR~\cite{Zatsarinny2005}, ROP~\cite{Marinkovic2019} and 206-state CCC~\cite{Fursa2005}.}
\end{figure*}

\begin{figure*}
\includegraphics[scale=0.92]{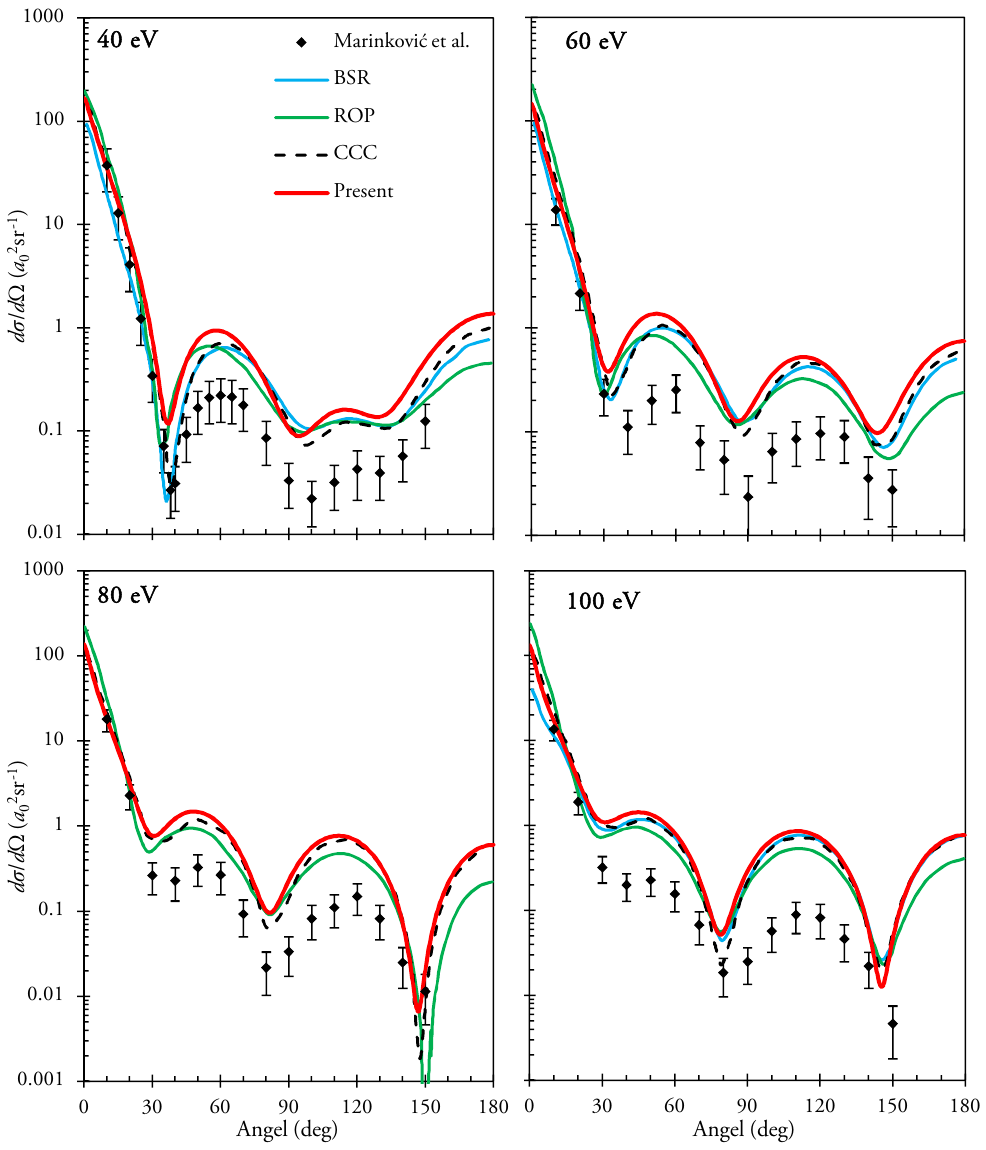}
\caption{\label{fig:zn2} Same as for figure~\ref{fig:zn1} but at 40, 60, 80 and 100 eV projectile energies.}
\end{figure*}

Our elastic differential cross section (DCS) values are shown in figures~\ref{fig:zn1} and ~\ref{fig:zn2}. They consist of comparisons with experiment and other theoretical approaches, including the relativistic optical potential (ROP) calculations of Marinkovi\'{c} et al.~\cite{Marinkovic2019}, the B-spline R-matrix (BSR) methodology of Zatsarinny et al.~\cite{Zatsarinny2005}, and the 206-state convergent close coupling (CCC) formulation of Fursa et al.~\cite{Fursa2005}. Our DCS results are in overall good agreement with other theoretical works for energies above 20 eV.
The agreement between theory and the experimental work of Marinkovi\'{c} et al.~\cite{Marinkovic2019} in all those energies is rather qualitative. This is quite interesting as the agreement between theoretical results for energies above 20 eV is rather remarkable. For energies $E \leq 20 \text{ eV}$, our DCS predictions do not inclusively agree with other theoretical results. Considerable disagreements exist on the locations of DCS minima with other calculations, except for 10 eV projectile energy, where our DCS values indicate the same minimum as that predicted by CCC calculations.

\begin{figure*}
\includegraphics{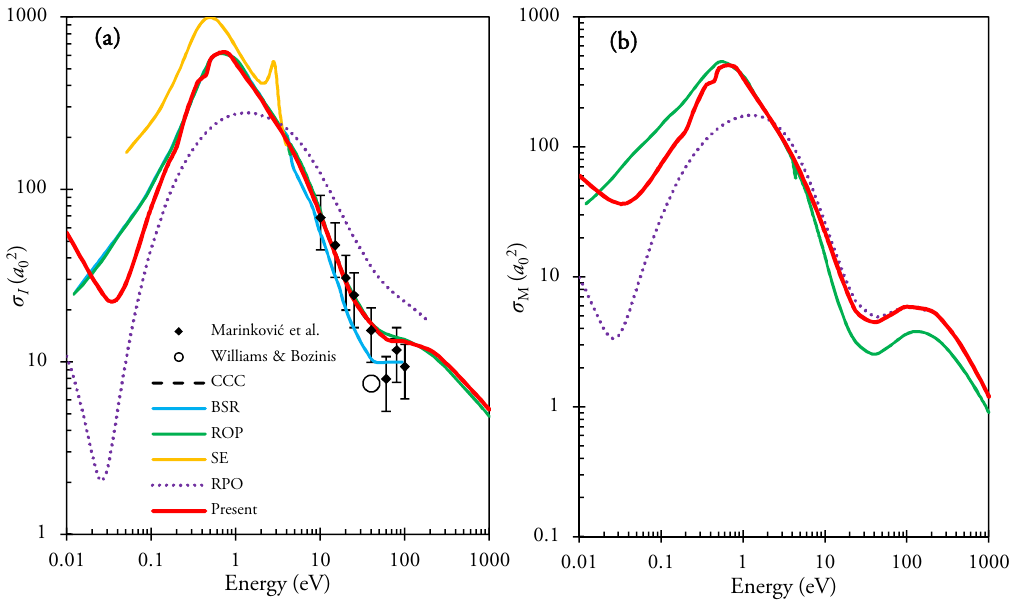}
\caption{\label{fig:zn3} Integrated (\textbf{a}) and momentum transfer (\textbf{b}) cross sections for elastic electron scattering from zinc: The legend in the figure describes markers for Present work; Experiment: Marinkovi\'{c} et al.~\cite{Marinkovic2019} and Williams and Bozinis~\cite{Williams1975}; Other theoretical: BSR~\cite{Zatsarinny2005}, ROP~\cite{Marinkovic2019}, 206-state CCC~\cite{Fursa2005}, SE~\cite{Arretche2022} and RPO~\cite{White2004}.}
\end{figure*}

Our elastic angle-integrated cross section (ICS), $\sigma_I$, and momentum transfer cross section (MTCS), $\sigma_M$, are presented in figure~\ref{fig:zn3}. 
Our predictions for ICS are in very good agreement with the experimental values of Marinkovi\'{c} et al.~\cite{Marinkovic2019} and CCC calculations for energies between 10 eV and 100  eV. Additionally, our ICS values are in good agreement with BSR and ROP values down to 0.2 eV, where our ICS curve dives toward a Ramsauer-Townsend (RT) minimum at about 0.06 eV. Our curve for $\sigma_I$ exhibits a maximum at the same energy as those of BSR and ROP calculations (about 0.7 eV). From right above the threshold for $(4s4p)^3P^0$ state ($\sim4$ eV) our ICS values do not follow, but stay higher than those of BSR calculations toward 100 eV. That is also the case for the values of the other theoretical calculations in this energy region. 

The ICS values by the semi-empirical (SE) calculations of Arretche et al.~\cite{Arretche2022}, at low energies, and the relativistic polarized-orbital (RPO) calculations of White et al.~\cite{White2004}, over a large interval, do not agree with other theoretical values as it can be seen in figure~\ref{fig:zn3}~a .

There are no experimental measurements for the zinc momentum transfer cross sections, which makes the comparisons purely theoretical. While in close agreement with the RPO values of White et al.~\cite{White2004} at high energies, our MTCS values sit higher than those by the ROP calculations down to about 7 eV. From that energy downward to the maximum value of MTCS, around 0.7 eV, our $\sigma_M$ values stay in a tight agreement with ROP’s. Further toward lower energies, our MTCS values present an RT minimum at about 0.06 eV, which is a different behavior than that shown by ROP values. The  agreement between our values and those of RPO calculations~\cite{White2004} also ends below 3 eV. Nonetheless, it is noted that RPO values also demonstrate an RT minimum at a slightly lower energy than that of the present work.

\begin{figure*}
\includegraphics{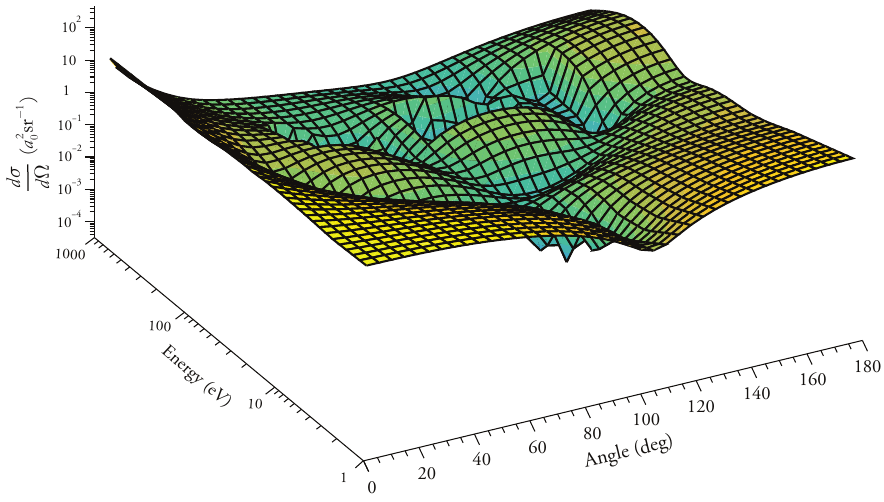}
\caption{\label{fig:zn4} A three-dimensional view of differential cross section for elastic electron scattering from zinc.}
\end{figure*}

To visualize the global behavior of the elastic differential cross section as a function of impact energy and scattering angle for elastic electron-zinc scattering, we present, in figure~\ref{fig:zn4}, a three-dimensional (3D) graph of DCS versus energy and the scattering angle.

\subsection{Cadmium}
\begin{figure*}
\includegraphics[scale=0.92]{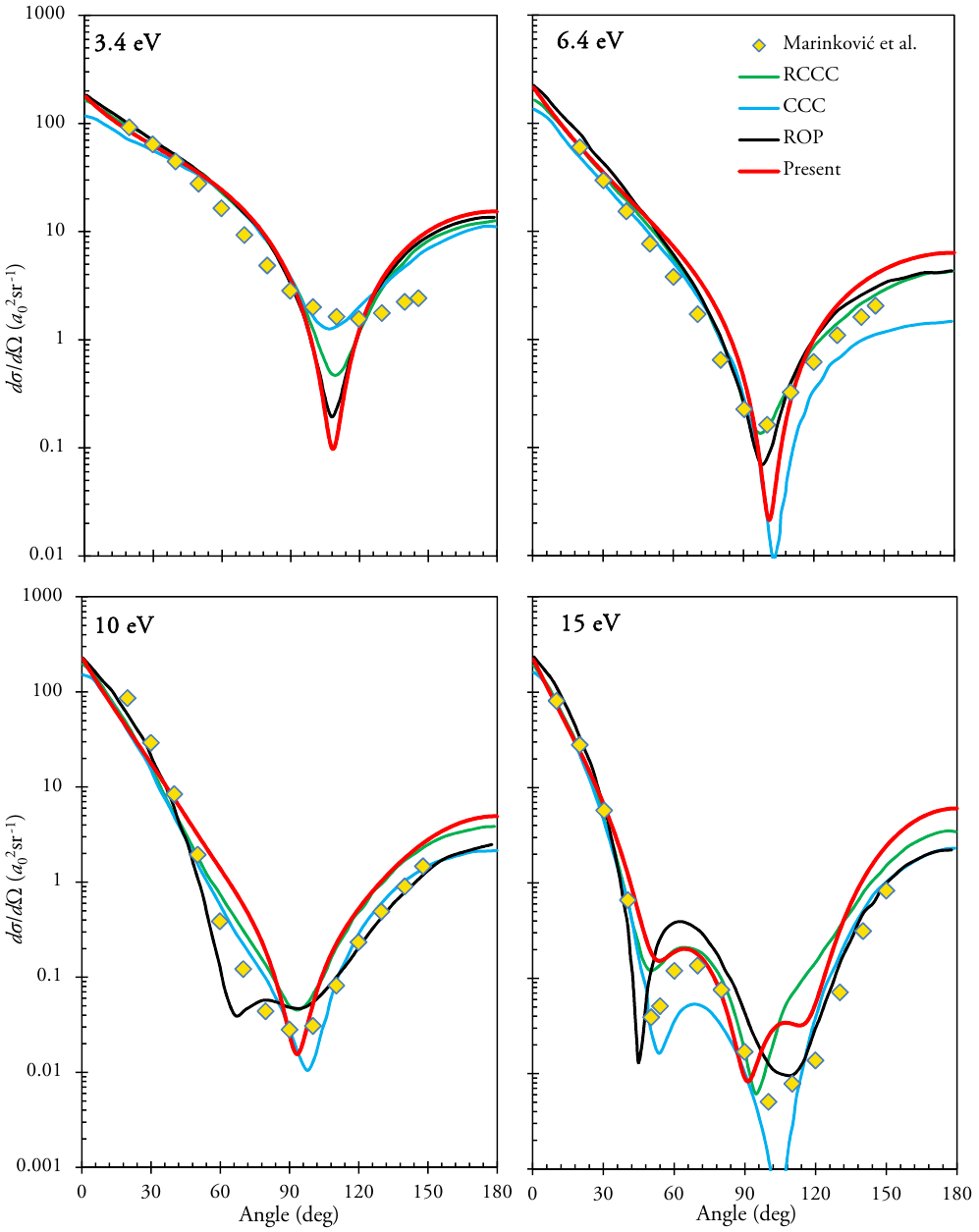}
\caption{\label{fig:cd1} Differential cross sections for elastic electron scattering from cadmium at 3.4, 6.4, 10 and 15 eV: The legend in the figure describes markers for Present work; Experiment: Marinkovi\'{c} et al.~\cite{Marinkovic2019}; Other theoretical: 200-state RCCC, ROP and 183-state CCC calculations of Ref.~\cite{Berrington2012}.}
\end{figure*}

\begin{figure*}
\includegraphics[scale=0.92]{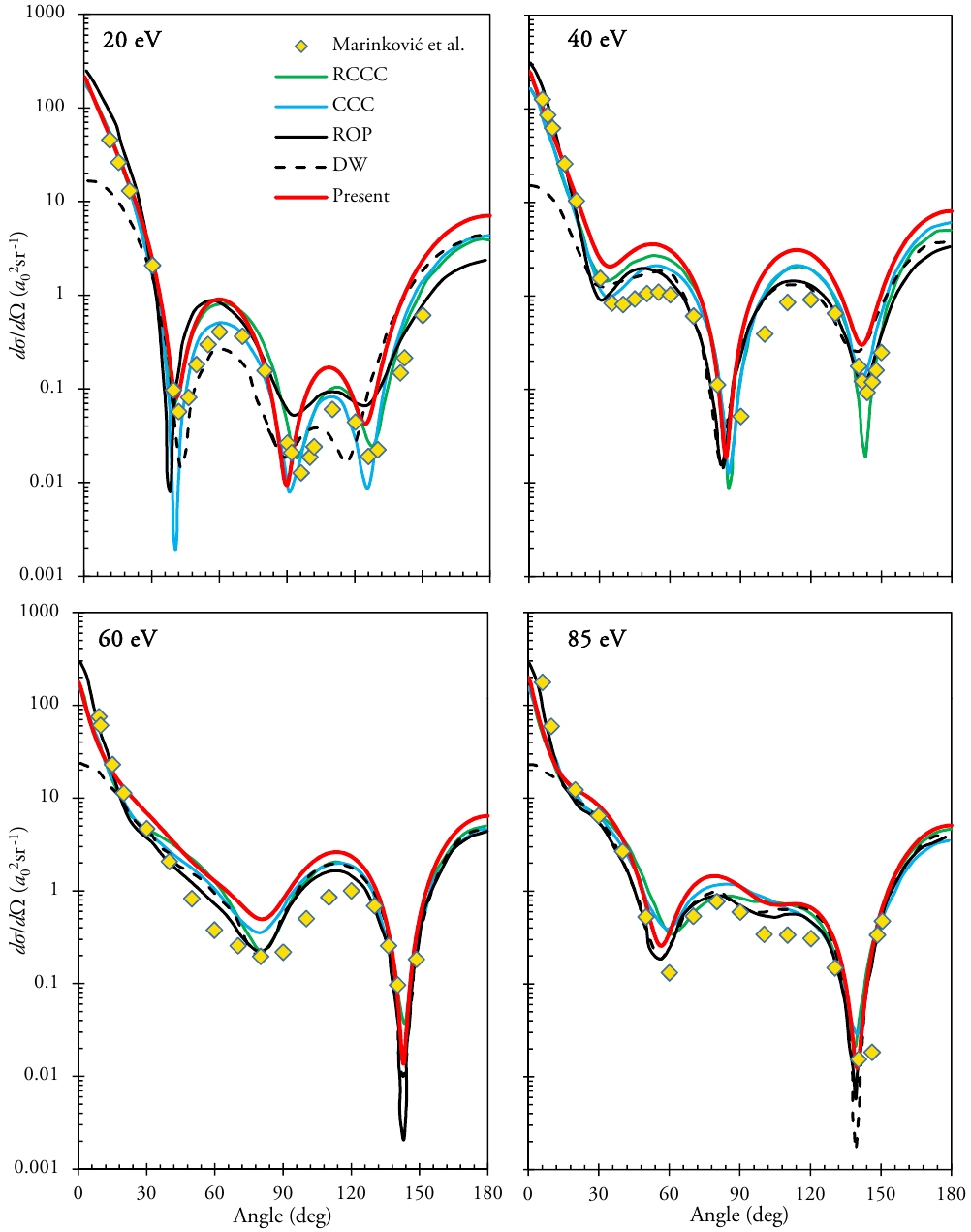}
\caption{\label{fig:cd2} Same as for figure~\ref{fig:cd1} but at 20, 40, 60 and 85 eV projectile energies. Additional theoretical work is DW calculations of Ref.~\cite{Madison1991}.}
\end{figure*}

\begin{figure*}
\includegraphics{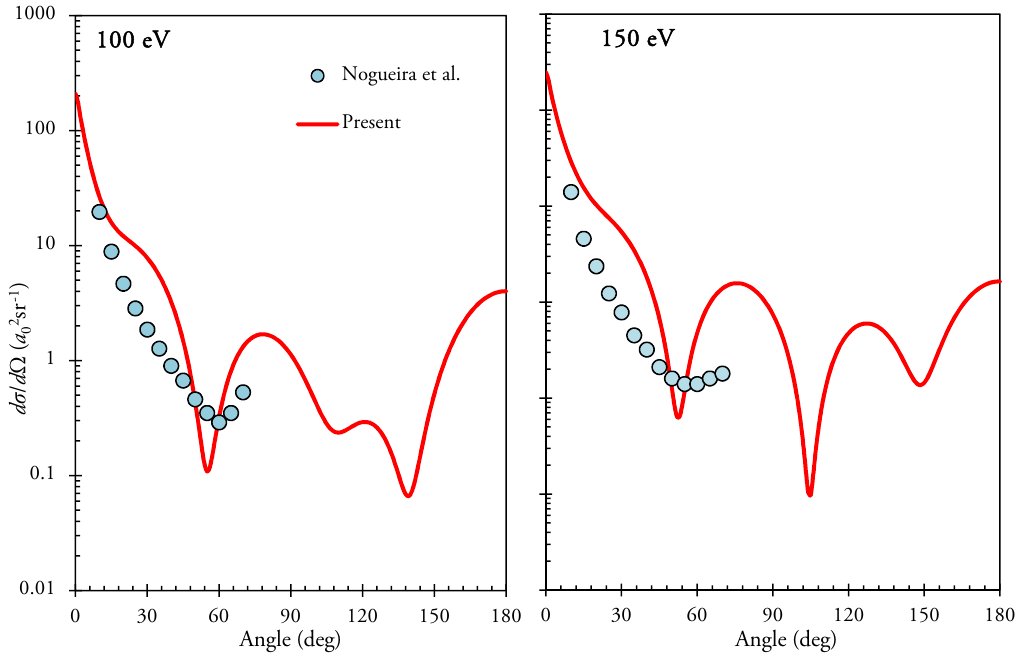}
\caption{\label{fig:cd3} Same as for figure~\ref{fig:cd2} but at 100 and 150 eV projectile energies with the sole experimental data of Nogueira et al.~\cite{Nogueira1987}.}
\end{figure*}

Present DCS results are shown in figures~\ref{fig:cd1}, \ref{fig:cd2} and ~\ref{fig:cd3}. These figures display comparisons with experimental values of  Marinkovi\'{c} et al.~\cite{Marinkovic1991} and Nogueira et al.~\cite{Nogueira1987}. Other theoretical investigations, in those figures, include a 200-state relativistic convergent close coupling (RCCC), a 183-state CCC and an ROP calculations of Berrington et al.~\cite{Berrington2012} and the semi-relativistic distorted-wave approach of Madison et al.~\cite{Madison1991}.

Our differential cross sections are in overall good agreement with the experimental data of Marinkovi\'{c} et al.~\cite{Marinkovic1991} except for 3.4 eV, at which energy all other theoretical results display the same disagreement with the experiment in terms of the depth of the minimum. Additionally, our theoretical DCS values show an interesting conformity at all impact energies except, maybe, for 15 eV.

Present DCS results at 100 and 150 eV are compared with the experimental data of Nogueira et al.~\cite{Nogueira1987} in figure~\ref{fig:cd3}. Within its limited angular range, the experiment displays pronounced differences with our DCS values. There are no other theoretical data at these two energies for comparison; however, we expect that our theoretical predictions are accurate.

\begin{figure*}
\includegraphics{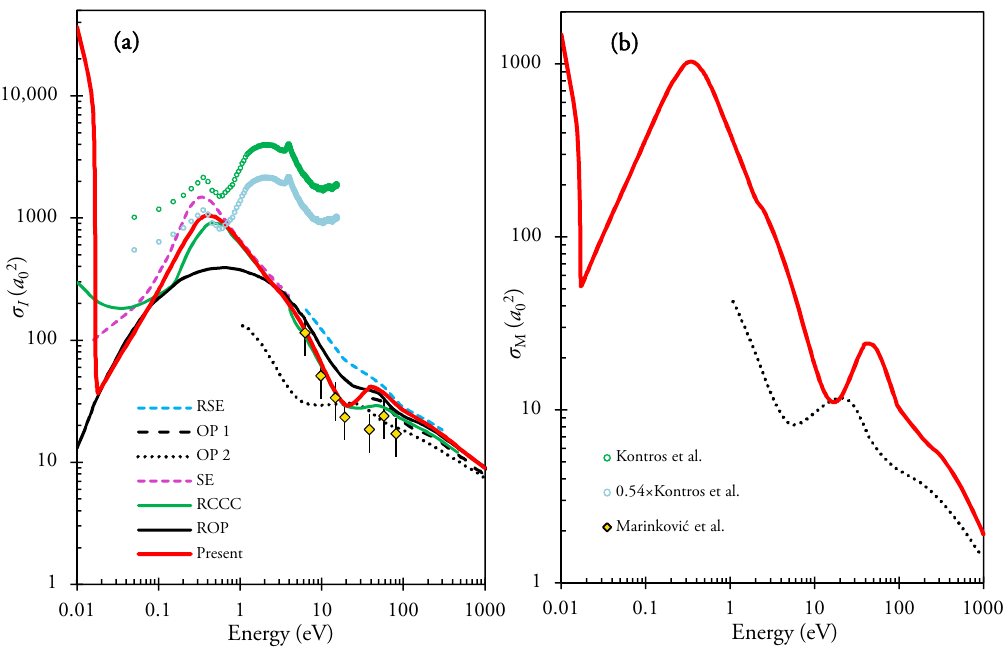}
\caption{\label{fig:cd4} Integrated (\textbf{a}) and momentum transfer (\textbf{b}) cross sections for elastic electron scattering from cadmium: The legends in the figure describe markers for Present work; Experiment: Nogueira et al.~\cite{Nogueira1987}, Kontros et al.~\cite{Kontros2002} and Marinkovi\'{c} et al.~\cite{Marinkovic2023}; Other theoretical: RSE~\cite{Nahar1991}, SE~\cite{Arretche2022}, OP 1~\cite{McGarrah1991}, OP 2~\cite{Haque2021}, RCCC~\cite{Marinkovic2023} and ROP~\cite{Marinkovic2023}.}
\end{figure*}

Our ICS and MTCS curves for energies between 0.01 and 1000 eV are presented in figure~\ref{fig:cd4} a and b, respectively. 
The experimental ICS data by Kontros et al.~\cite{Kontros2002} within the energy range of 0.05 to 4 eV indicate a local ICS maximum at around 0.45 eV. Our ICS values exhibit an absolute maximum at 0.4 eV, which is not far from the experimental value. We took the liberty to renormalize the experimental data of Kontros et al.~\cite{Kontros2002} by a multiplicative factor of 0.54. The agreement below 0.5 eV is acceptable; above that value the experimental data indicate the opening of other excitation channels before reaching the ionization limit. The recent measurements of Marinkovi\'{c} et al.~\cite{Marinkovic2023} are in reasonably good agreement with our calculations. Agreement between our ICS values and those of other theoretical works is, however, varying. The agreement with the optical potential calculations of McGarrah et al.~\cite{McGarrah1991} and Haque et al.~\cite{Haque2021} is qualitative, down to 30 eV. Although our ICS values are in excellent agreement with those of the relativistic semi-empirical (RSE) approach of Nahar~\cite{Nahar1991} from 200 down to 30 eV, for energies lower than 30 eV there is considerable disagreement. The ICS values of the semi-empirical (SE) calculations of Arretche et al.~\cite{Arretche2022} are in excellent agreement with our values down to about 1 eV, but below 1 eV they only display a comparable behavior to that of our ICS curve down to 0.08 eV.  The RCCC values of Marinkovi\'{c} et al.~\cite{Marinkovic2023} are in overall good agreement with our calculations. The RCCC results are lower than ours for energies above 20 eV displaying a better agreement with experiment. Similar to zinc, our ICS values indicate an RT minimum at about 0.018 eV.  Only the RCCC calculations exhibit a minimum in that area, albeit a broader one. It is worth noting that the energy ranges are extremely narrow in this region.

Similar to zinc, there are no experimental measurements for the cadmium momentum transfer cross sections and only one set of theoretical data, that of the optical potential (OP 2) calculations of Haque et al.~\cite{Haque2021}. The comparison shows, at best, a qualitative agreement between the present MTCS values and those of Haque et al.~\cite{Haque2021}. Our MTCS curve also displays an RT minimum at about 0.018 eV, where there are no other data to be compared with.

A 3D graph of DCS versus energy and the scattering angle for elastic electron-cadmium scattering is presented in figure~\ref{fig:cd4}. Compared to that of zinc, in figure~\ref{fig:zn4}, cadmium’s 3D graph presents a more complex atomic structure, as expected.

\begin{figure*}
\includegraphics{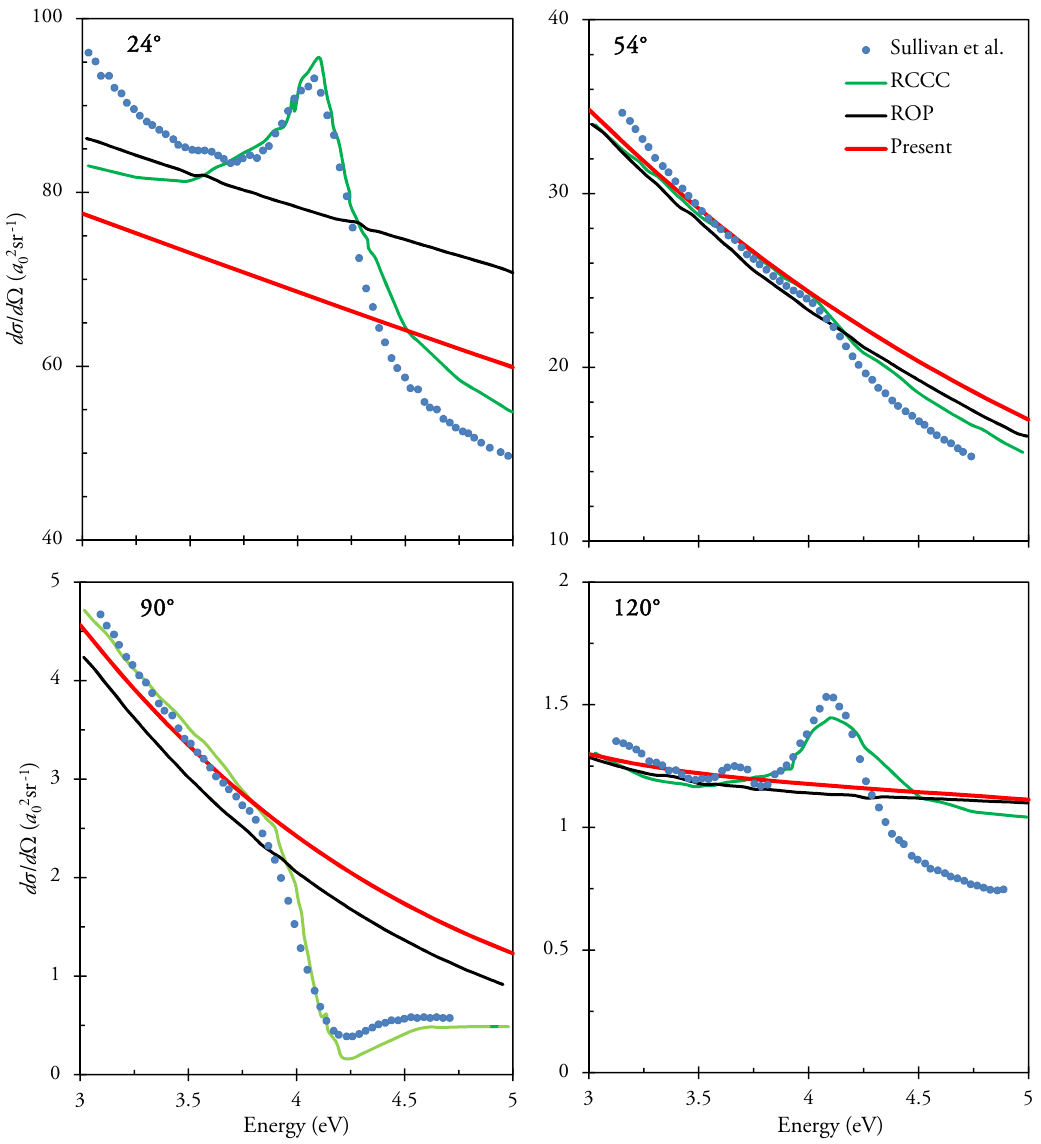}
\caption{\label{fig:cd6} Differential cross sections for elastic electron scattering from cadmium at scattering angles $24^\circ$, $54^\circ$, $90^\circ$ and $120^\circ$ versus projectile energy: The legend in the figure describes markers for Present work; Experiment: Sullivan et al.~\cite{Sullivan2003}; Other theoretical: RCCC and ROP calculations of Ref.~\cite{Berrington2012}.}
\end{figure*}

Finally, in figure~\ref{fig:cd6}, we compare our DCS values at low-energy elastic electron-cadmium scattering at various scattering angles with those of experimental and other theoretical results. The purpose of obtaining these cross sections shown in the graphs was to demonstrate the existence of a low energy resonance structure and its subsequent effect on elastic $e$-Cd scattering. Clearly, the 55-state RCCC calculations of Berrington et al.~\cite{Berrington2012} confirms the existence of such resonance structures with very good agreements with the experimental data of Sullivan et al.~\cite{Sullivan2003}. Our DCS predictions and those of the ROP method~\cite{Berrington2012} show no sign of resonance in the elastic scattering spectrum, which was expected from a single-channel approach.

\subsection{Mercury}
\begin{figure*}
\includegraphics[scale=0.92]{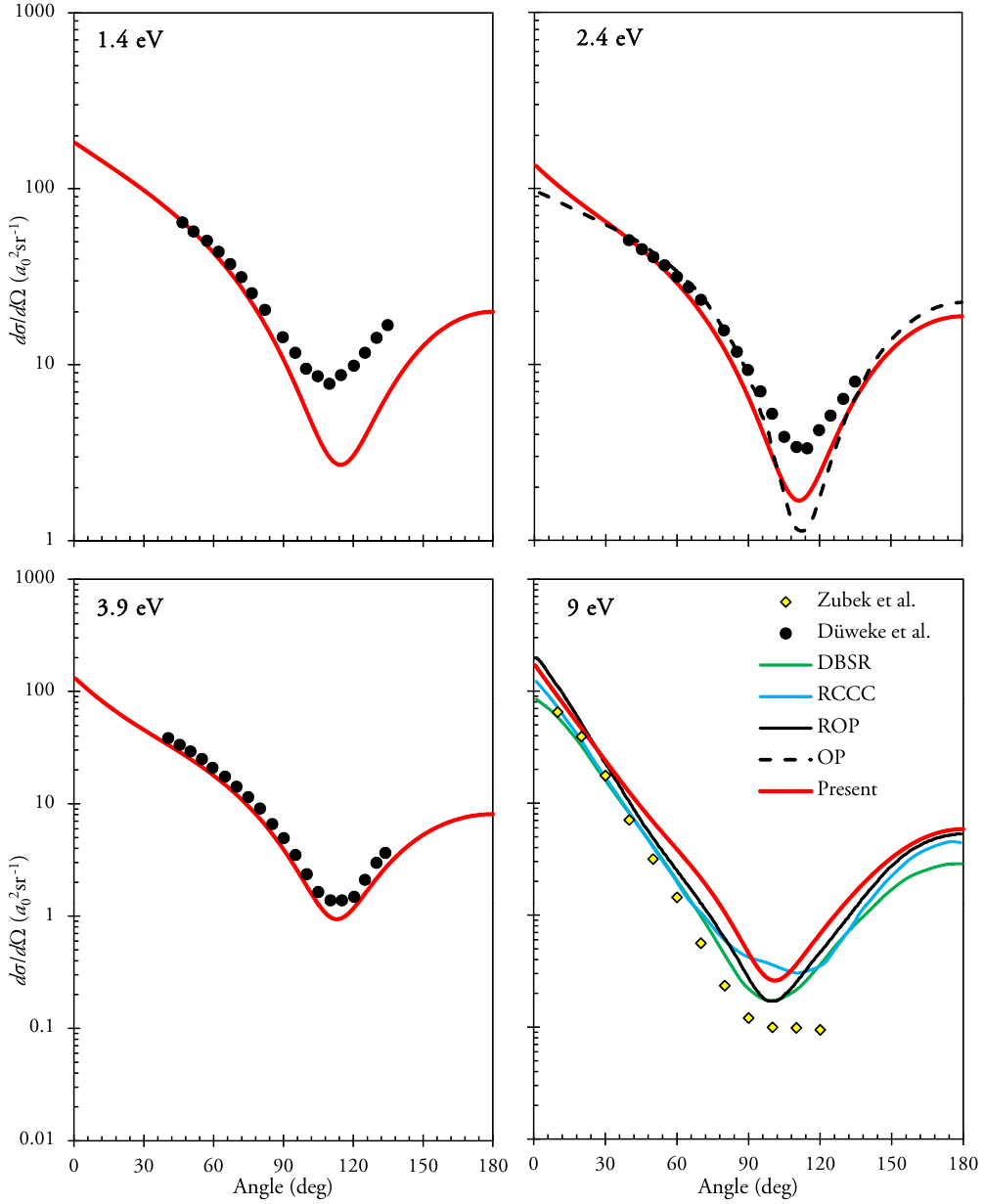}
\caption{\label{fig:hg1} Differential cross sections for elastic electron scattering from mercury at 1.4, 2.4, 3.9 and 9 eV: The legend in the figure describes markers for Present work; Experiment: Zubek et al.~\cite{Zubek1995} and D\"{u}weke et al.~\cite{Duweke1976}; Other theoretical: 36-state DBSR~\cite{Zatsarinny2009}, 193-state RCCC~\cite{Bostock2010}, ROP~\cite{Sienkiewicz1997} and OP~\cite{Haque2021} calculations.}
\end{figure*}

\begin{figure*}
\includegraphics[scale=0.92]{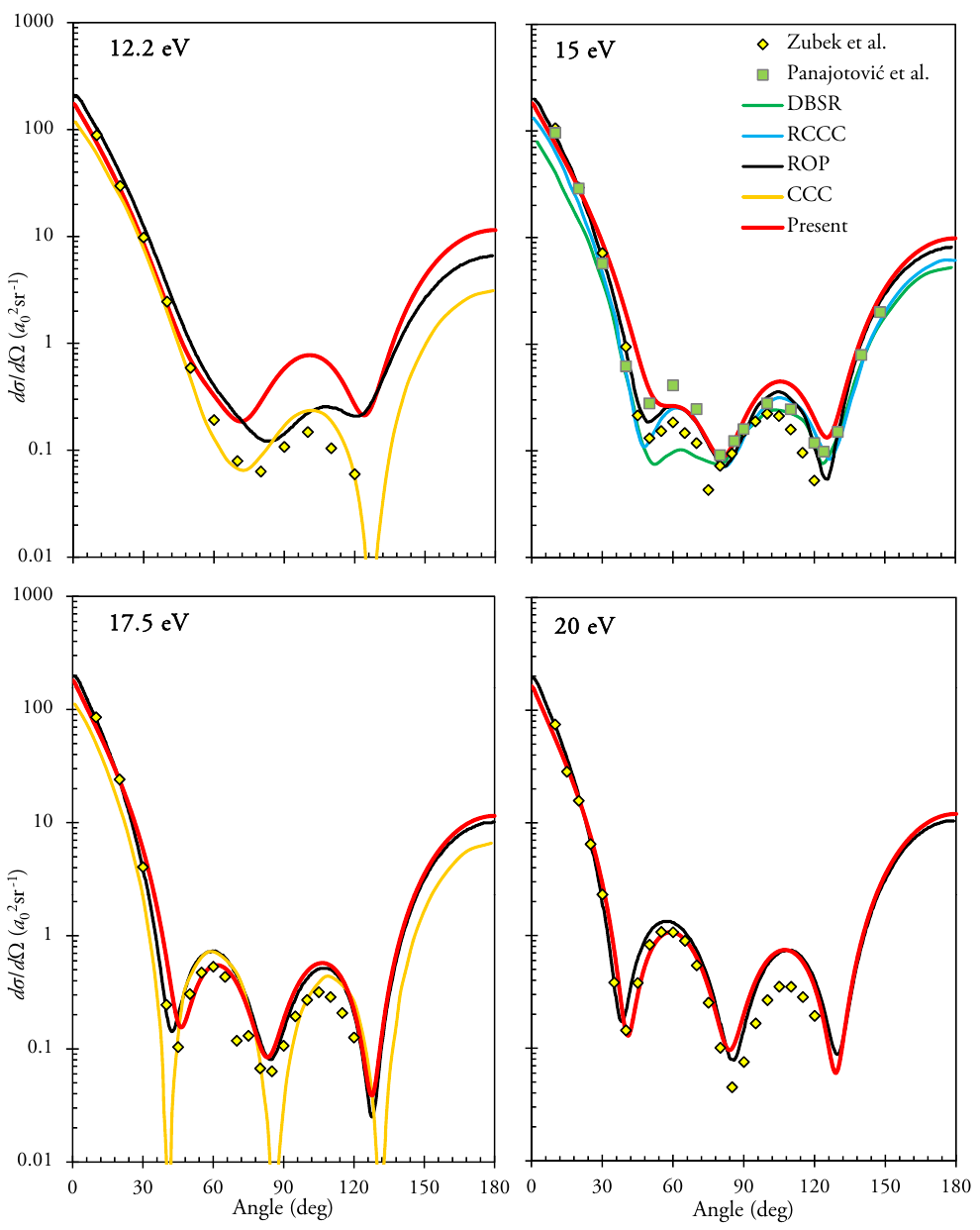}
\caption{\label{fig:hg2} Differential cross sections for elastic electron scattering from mercury at 12.2, 15, 17.5 and 20 eV: The legend in the figure describes markers for Present work; Experiment: Zubek et al.~\cite{Zubek1995} and Panajotovi\'{c} et al.~\cite{Panajotovic1993}; Other theoretical: 36-state DBSR~\cite{Zatsarinny2009}, 193-state RCCC~\cite{Bostock2010}, ROP~\cite{Sienkiewicz1997} and 54-state CCC~\cite{Fursa2003a} calculations.}
\end{figure*}

\begin{figure*}
\includegraphics[scale=0.92]{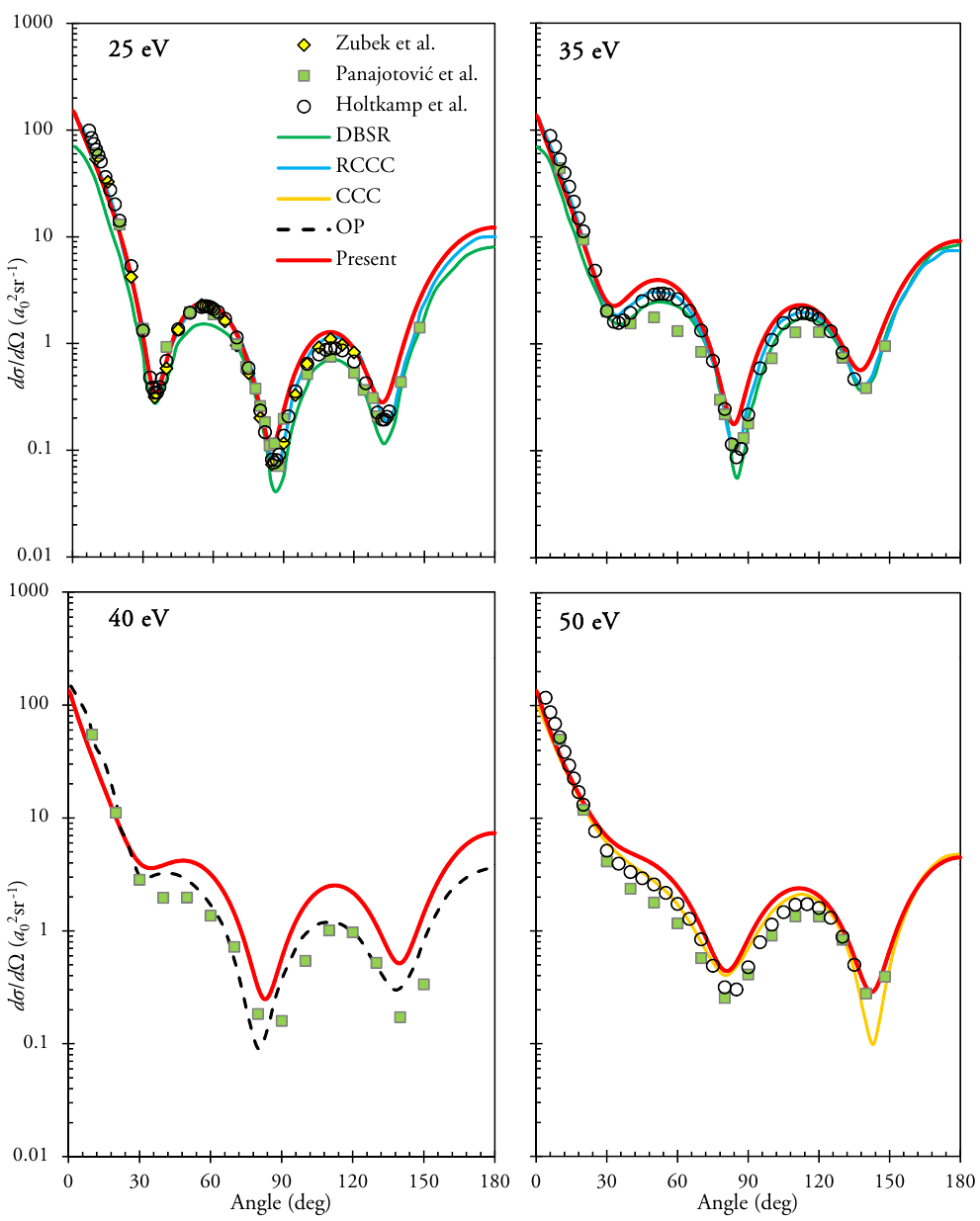}
\caption{\label{fig:hg3} Differential cross sections for elastic electron scattering from mercury at 25, 35, 40 and 50 eV: The legend in the figure describes markers for Present work; Experiment: Zubek et al.~\cite{Zubek1995}, Panajotovi\'{c} et al.~\cite{Panajotovic1993} and Holtkamp et al.~\cite{Holtkamp1987} for which the size of the marker demonstrates the error; Other theoretical: 36-state DBSR~\cite{Zatsarinny2009}, 193-state RCCC~\cite{Bostock2010}, 54-state CCC~\cite{Fursa2003a} and OP~\cite{Haque2021} calculations.}
\end{figure*}

\begin{figure*}
\includegraphics[scale=0.92]{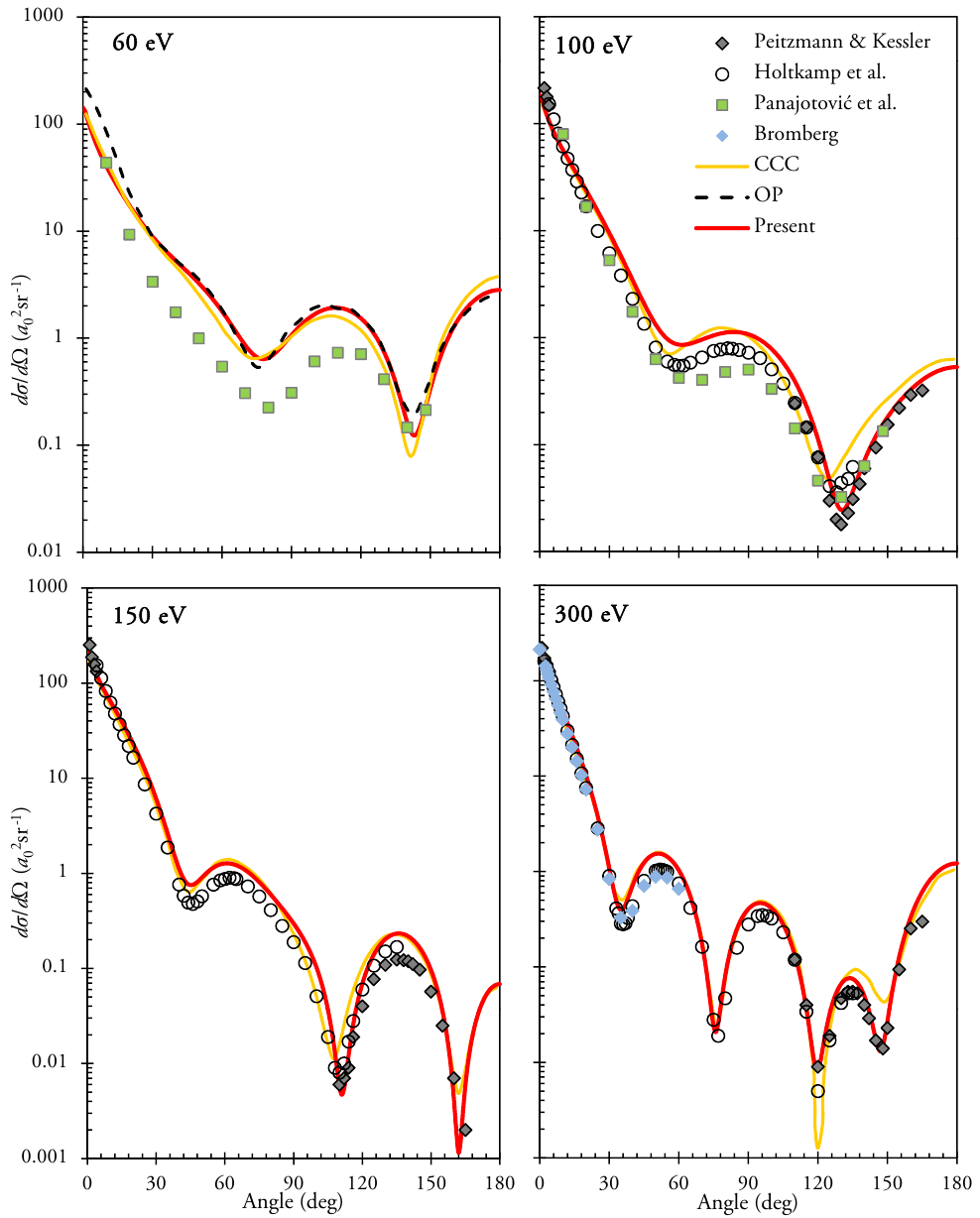}
\caption{\label{fig:hg4} Differential cross sections for elastic electron scattering from mercury at 60, 100, 150 and 300 eV: The legend in the figure describes markers for Present work; Experiment: Panajotovi\'{c} et al.~\cite{Panajotovic1993}, Holtkamp et al.~\cite{Holtkamp1987}, Peitzmann and Kessler~\cite{Peitzmann1990} and Bromberg~\cite{Bromberg1969}; Other theoretical: 54-state CCC~\cite{Fursa2003a} and OP~\cite{Haque2021} calculations.}
\end{figure*}

\begin{figure*}
\includegraphics[scale=0.92]{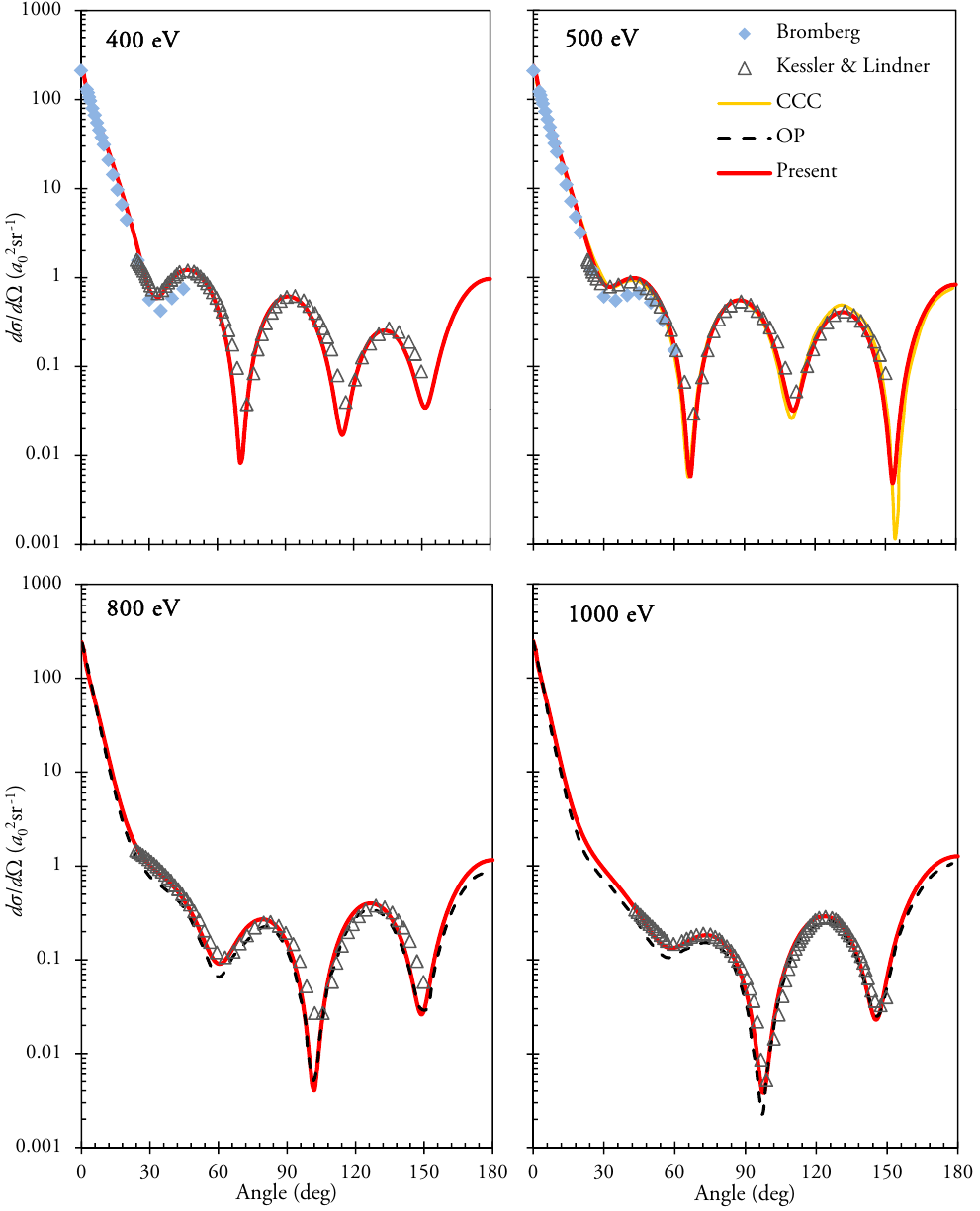}
\caption{\label{fig:hg5} Differential cross sections for elastic electron scattering from mercury at 400, 500, 800 and 1000 eV: The legend in the figure describes markers for Present work; Experiment: Bromberg~\cite{Bromberg1969} and Kessler and Lindner~\cite{Kessler1965}; Other theoretical: 54-state CCC~\cite{Fursa2003a} and OP~\cite{Haque2021} calculations.}
\end{figure*}

We present our DCS values for elastic electron scattering off mercury in figures~\ref{fig:hg1}~ -- ~\ref{fig:hg5}. Comparisons are made with the experimental works of Zubek et al.~\cite{Zubek1995}, D\"{u}weke et al.~\cite{Duweke1976},  Panajotovi\'{c} et al.~\cite{Panajotovic1993}, Holtkamp et al.~\cite{Holtkamp1987}, Bromberg~\cite{Bromberg1969}, Peitzmann and Kessler~\cite{Peitzmann1990} and Kessler and Lindner~\cite{Kessler1965} as well as a collection of theoretical works, which includes the 36-state Dirac B-spline R-matrix (DBSR) calculations of Zatsarinny and Bartschat~\cite{Zatsarinny2009}, the 193-state RCCC investigations of Bostock et al.~\cite{Bostock2010}, the 54-state CCC formulation of Fursa et al.~\cite{Fursa2003a}, the frozen-core Dirac-Fock potential and a polarization model potential (ROP) calculations of Sienkiewicz~\cite{Sienkiewicz1997}, and the Dirac relativistic partial wave analysis with a complex projectile-atom optical potential (OP) investigations of Haque et al.~\cite{Haque2021}.

Our differential cross sections are in excellent agreement with the measurements of D\"{u}weke et al.~\cite{Duweke1976} at 3.9 eV, while at 1.4 and 2.4 eV our DCS minimum is more pronounced. At 9 eV, while the displayed calculations are more or less in agreement on the position of DCS minimum, none (including ours) is in an acceptable agreement with the experimental data by Zubek et al.~\cite{Zubek1995} beyond the forward direction.

At 12.2 eV impact energy, our DCS values agree with the experimental data~\cite{Zubek1995} only in the forward scattering angles, while CCC predictions are in excellent agreement with experiment. 

The present DCS curve at 15 eV displays three minima but overall favors the measurements of Panajotovi\'{c} et al.~\cite{Panajotovic1993} over those of Zubek et al.~\cite{Zubek1995}. Above 15 eV and through 25 eV, our differential cross sections display very good agreement with the experimental data of Zubek et al.~\cite{Zubek1995}.

At 17.5 eV, our values slightly disagree with the ROP~\cite{Sienkiewicz1997} and CCC~\cite{Fursa2003a} results on the positions of the minima, and on their depths with the CCC~\cite{Fursa2003a} predictions. The same disagreements on the positions of the minima are observed for 20 eV impact energy with the ROP~\cite{Sienkiewicz1997}, where our DCS values are in excellent agreement with experiment in the forward scattering angles up to 90 degrees.

At 25 and 35 eV, single and multi channel calculations are in very good agreement with each other and the experimental data. That means the quality of the scattering potential is the deterministic factor in cross section calculations at such energies. Marked discrepancies are observed in comparisons with experimental data of Panajotovi\'{c} et al.~\cite{Panajotovic1993} at 40 and 60 eV. There is, however, good agreement at 50 and 100 eV with the measurements of Panajotovi\'{c} et al.~\cite{Panajotovic1993}. At energies higher than 100 eV, our DCS predictions are in excellent agreement with various experimental data and, to some extent, with other theoretical works. These agreements include the measurements of Holtkamp et al.~\cite{Holtkamp1987}, available between 25 eV and 300 eV, which are in excellent agreement with various theories, and the experimental data of Peitzmann and Kessler~\cite{Peitzmann1990}, for 100 eV and 150 eV, as well as those of Bromberg~\cite{Bromberg1969} for 300 eV. At 400, 500, 800, and 1000 eV the experimental data of Bromberg~\cite{Bromberg1969} and Kessler and Lindner~\cite{Kessler1965} agree well with our predictions as well as those of CCC~\cite{Fursa2003a} and OP~\cite{Haque2021} calculations.
Altogether, previous theoretical results are not available below 9 eV impact energy, above which they agree remarkably well, except for the CCC values that display deeper minima.

\begin{figure*}
\includegraphics{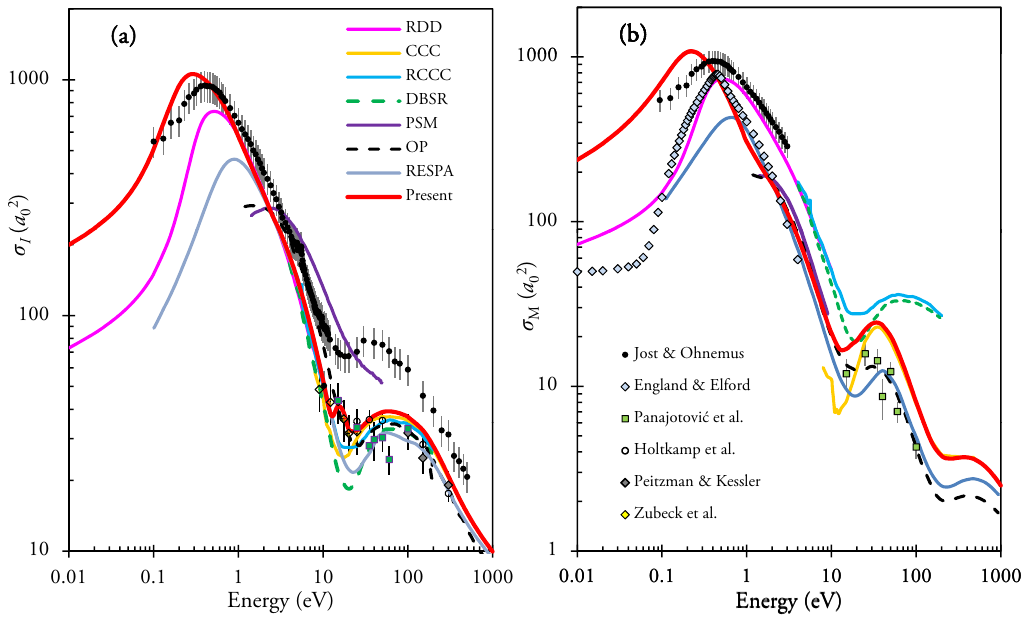}
\caption{\label{fig:hg6} Integrated (\textbf{a}) and momentum transfer (\textbf{b}) cross sections for elastic electron scattering from mercury: The legends in the figure describe markers for Present work; Experiment: Zubek et al.~\cite{Zubek1995}, Panajotovi\'{c} et al.~\cite{Panajotovic1993} and Holtkamp et al.~\cite{Holtkamp1987}, Peitzmann and Kessler~\cite{Peitzmann1990}, Jost and Ohnemus~\cite{Jost1979} and England and Elford~\cite{England1991}. Other theoretical: 36-state DBSR~\cite{Zatsarinny2009}, 193-state RCCC~\cite{Bostock2010}, 54-state CCC~\cite{Fursa2003a}, OP~\cite{Haque2021}. PSM~\cite{McEachran1987} and RDD~\cite{McEachran2003} calculations.}
\end{figure*}

In figures~\ref{fig:hg6} a and b, ICS and MTCS comparisons of our values with experiment and other theoretical works for elastic electron scattering from mercury are displayed for energies between 0.01 and 1000 eV. These comparisons include a number of aforementioned works. They also include measurements by Jost and Ohnemus~\cite{Jost1979} and England and Elford~\cite{England1991}, as well as, the potential scattering model (PSM) calculations of McEachran and Stauffer~\cite{McEachran1987} and the relativistic dynamic distortion (RDD) calculations of McEachran et al.~\cite{McEachran2003}.

The experimental ICS data by Zubek et al.~\cite{Zubek1995}, Holtkamp et al.~\cite{Holtkamp1987}, and Peitzmann and Kessler~\cite{Peitzmann1990} show close agreement where they overlap within the energy range of 10 to 300 eV. 

Our ICS values display excellent agreement with the values from experiment and convergent close coupling calculations~\cite{Fursa2003a,Bostock2010} down to low impact energies. At energies between 10 and 25 eV our ICS curve closely follows the experimental data of Zubek et al.~\cite{Zubek1995}, while disagreeing with other theoretical values, e.g., the DBSR~\cite{Zatsarinny2009} values which exhibit a deeper minimum in this energy region compared to others.
For energies below 10 eV, our ICS values are in excellent agreement with the DBSR~\cite{Zatsarinny2009} and RCCC~\cite{Bostock2010} values. The experimental data of Jost and Ohnemus~\cite{Jost1979} agree well with our calculations up to about 5 eV. Thereafter, the experimental values are higher since they include the onset of inelastic channels.

\begin{figure*}
\includegraphics{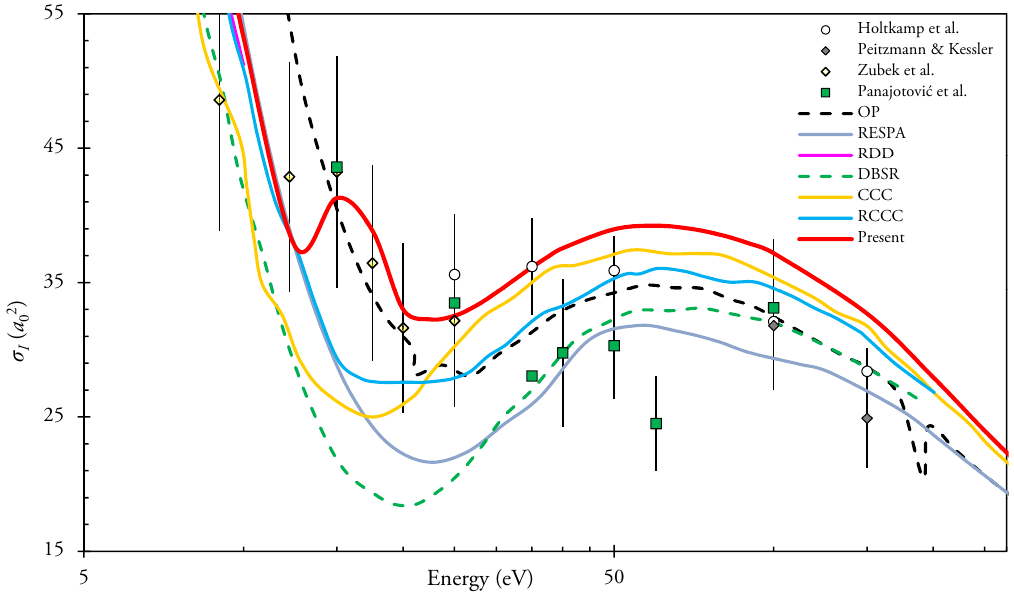}
\caption{\label{fig:hg7} A magnified view of the integrated cross section plot of figure \ref{fig:hg6} for the impact energies between 5 and 275 eV. The same caption as that of figure \ref{fig:hg6} applies.}
\end{figure*}

To emphasize the degree of agreement between various theories and experiments we show in figure~\ref{fig:hg7} the enlarged portion of the ICS graph for energies between 5 and 275 eV. Of all calculations, ours uniquely predicts a structure around 15 eV, which seems to be corroborated by the data of Zubek et al.~\cite{Zubek1995}.

Unlike zinc and cadmium, for the mercury momentum transfer cross sections there are three experimental measurements, i.e., Refs.~\cite{Jost1979,England1991,Panajotovic1993}, as well as six other theoretical treatments, i.e., Refs.~\cite{McEachran1987, McEachran2003,Fursa2003a,Zatsarinny2009,Bostock2010,Haque2021}.  Our data exhibit a ``behavioral'' agreement with measurements of Panajotovi\'{c} et al.~\cite{Panajotovic1993} at energies between 10 and 100 eV and some agreements with the experiment by Jost and Ohnemus~\cite{Jost1979} between 0.1 and 1 eV. The experimental data of England and Elford~\cite{England1991} while displaying a similar shape to that of Jost and Ohnemus~\cite{Jost1979} are in good agreement with our values down to 0.4 eV. The theoretical results of RDD method~\cite{McEachran2003} overlap fittingly with the experimental data of Jost and Ohnemus~\cite{Jost1979} down to 0.9 eV while sitting higher than our values.

\begin{figure*}
\includegraphics{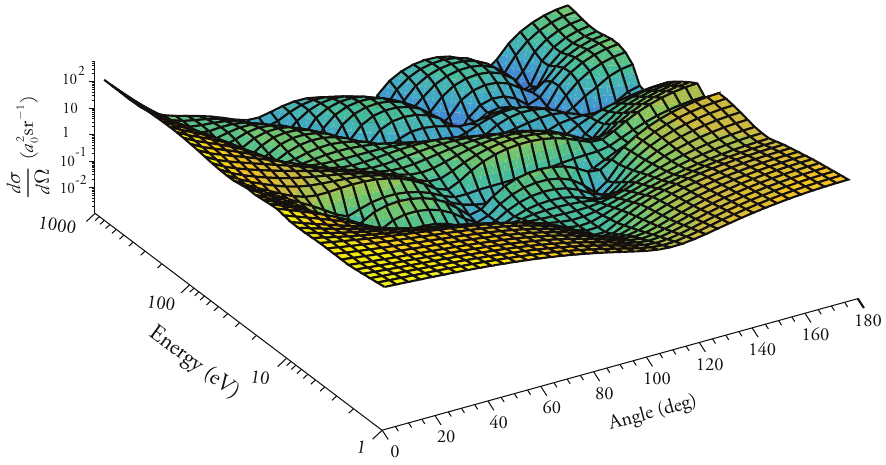}
\caption{\label{fig:hg8} A three-dimensional view of differential cross section for elastic electron scattering from mercury.}
\end{figure*}

Finally, we present a 3D graph of DCS versus energy and the scattering angle for elastic scattering of electrons off mercury in figure~\ref{fig:hg8}. Mercury's 3D depiction of elastic DCS displays more complexity compared to those of of zinc and cadmium, displayed in figures~\ref{fig:zn3} and \ref{fig:cd4}.

\section{Conclusion}
We now have extended our method of calculations for inert gases and alkaline-earth-metal atoms to Zn, Cd, and Hg with outer shell configuration $nd^{10}(n+1)s^2$. Our method is relatively simple: a self-consistent calculation of the target potential plus a core-polarization potential with only one adjustable parameter, which is the cut-off radius of the polarization potential. Our DCS values compare well with the available experimental DCS data and other sophisticated calculations.  The present ICS values also reached good agreement with the results of other calculations.  Comparisons between our cross sections with prior theoretical results generated both agreements and disagreements. The latter were present when the effect of virtual excitation of core states was important.  This was particularly the case for cadmium. We believe our present extensive study, together with the other dependable calculations, referred to in the current paper, can serve in identifying the experimental data that are the most accurate.  In summary, we expanded the available data on elastic electron scattering from zinc, cadmium and mercury to serve as guidance to future relevant experimental efforts.

\bibliography{zcmcs}

\begin{figure*}
\includegraphics{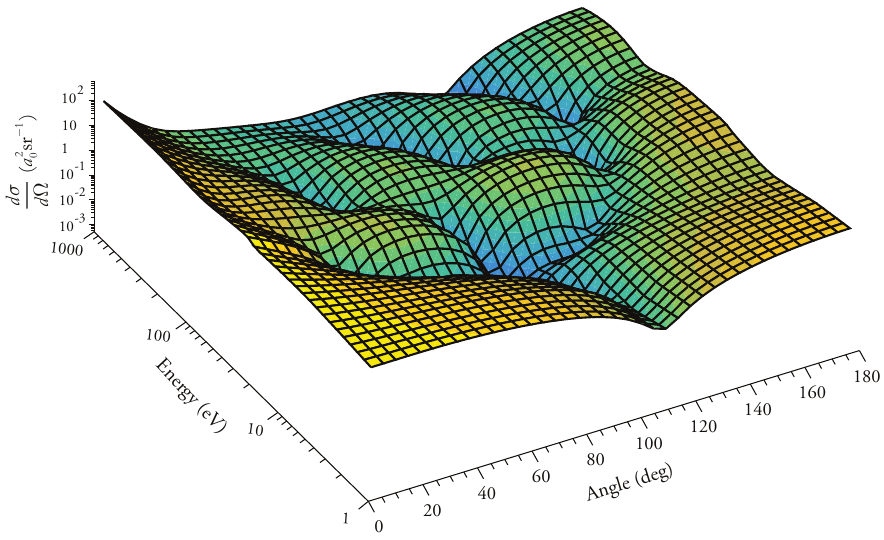}
\caption{\label{fig:cd5} A three-dimensional view of differential cross section for elastic electron scattering from cadmium.}
\end{figure*}

\end{document}